\documentclass[aps,
amsmath,amssymb,reprint,superscriptaddress]{revtex4-2}

\usepackage{graphicx}
\usepackage{dcolumn}
\usepackage{bm}
\usepackage[utf8]{inputenc}
\usepackage[T1]{fontenc}
\usepackage{booktabs, array, mathptmx, float, tabularx, booktabs, lipsum, amsmath, amssymb}
\usepackage{url}
\usepackage{caption, subcaption}
\usepackage{siunitx, xcolor}
\usepackage[version=4]{mhchem}
\usepackage[colorlinks,linkcolor=blue,anchorcolor=blue,citecolor=blue]{hyperref}

\begin{document}

\preprint{APS/123-QED}

\title{Switching between Skyrmions and Yoshimori Spin Spirals via Li Absorption in Janus Magnets}

\author{Xinyuan Jiang}
\affiliation{State Key Laboratory of Low Dimensional Quantum Physics, Department of Physics, Tsinghua University, Beijing 100084, China}
\author{Jian Wu}
\affiliation{State Key Laboratory of Low Dimensional Quantum Physics, Department of Physics, Tsinghua University, Beijing 100084, China}
\affiliation{Frontier Science Center for Quantum Information, Beijing, China}
\author{Weiyi Pan}
\email{weiyi.pan@physik.uni-regensburg.de}
\affiliation{State Key Laboratory of Low Dimensional Quantum Physics, Department of Physics, Tsinghua University, Beijing 100084, China}
\affiliation{Institute for Theoretical Physics, University of Regensburg, 93040 Regensburg, Germany}

\date{\today}

\begin{abstract}

Chiral magnetic textures have attracted considerable attention owing to their topological properties and potential applications in spintronic devices. Here, we employ first-principles calculations together with atomic spin dynamics simulations to explore the switching between skyrmions and Yoshimori-type spin spirals induced by Li adsorption in Janus two-dimensional (2D) CrTeSe. We show that selective Li adsorption on either the Se- or Te-terminated surface stabilizes distinct magnetic phases: Li adsorption on the Se side favors a Yoshimori-type spin spiral, whereas adsorption on the Te side stabilizes the skyrmionic state. This contrast originates from site-dependent modifications of exchange interactions, magnetic anisotropy (MA), and the Dzyaloshinskii–Moriya interaction (DMI). In addition, the response of magnetic textures to out-of-plane magnetic fields differs strongly between the two systems. These results demonstrate that surface adsorption provides an effective strategy for reversible control of chiral magnetic states in 2D magnets, while also offering fundamental insights into the competing interactions that govern the stability of skyrmions and Yoshimori spin spirals. Our findings highlight the potential of Janus 2D materials as a versatile platform for engineering tunable spintronic devices.
\end{abstract}

%\keywords{}

\maketitle

\section{\label{sec:level1}INTRODUCTION}

Chiral magnetic structures, including spin spirals \cite{Bode2007, Franken2014, Ryu2013}, skyrmions \cite{doi:10.1126/science.1166767, Yu2010, Fert2017} and bimerons \cite{Gao2019, PhysRevApplied.12.064054, Li2020}, have emerged as a central focus in condensed matter physics and spintronics due to their unique topological properties and potential for device applications. These nontrivial spin textures exhibit remarkable properties: Skyrmions, for example, possess topological protection, which endows them with enhanced stability and the potential to resist defects and thermal fluctuations \cite{Cortés-Ortuño2017, 10.1063/5.0009559}, while spin-spiral magnets features topological magnetism \cite{Fert2017}, multiferroicity \cite{Song2022}, and chirality-dependent transport behaviors \cite{PhysRevResearch.3.043155}. Their small size and lower threshold current densities make them promising building blocks for high-density, energy-efficient spintronic devices such as racetrack memories, logic gates, and magnetic sensors \cite{PhysRevApplied.12.034005, Wang2020}. However, practical integration requires a fundamental understanding of the microscopic interactions that govern their formation, stability and dynamics. Such insight is essential for achieving controlled manipulation of chiral magnetic states and enabling the design of next-generation spintronic devices.

The formation of chiral magnetic structures typically arises from the interplay of several competing magnetic interactions \cite{PhysRevB.105.214435, PhysRevB.109.024420,PhysRevB.111.134421}. Among them, the Dzyaloshinskii-Moriya interaction (DMI) \cite{PhysRevLett.4.228, DZYALOSHINSKY1958241, PhysRev.120.91} originating from spin-orbit coupling (SOC) in systems lacking inversion symmetry is recognized as a key driving force for the chiral magnetic order. The DMI favors noncollinear spin configurations, with the direction of its vector determining the chirality and spatial profile of the resulting magnetic texture \cite{PhysRevLett.115.267210, Yang2018, Yang2023}. In two-dimensional (2D) materials with broken inversion symmetry, such as Janus materials and 2D heterostructures, the DMI can stabilize chiral helical spin textures \cite{PhysRevB.102.014451, PhysRevB.105.224403, D3CP02470A}. In addition to DMI, exchange frustration \cite{D3CP02470A, Hu2017, vonMalottki2017}, which arises from the competition between short-range ferromagnetic (FM) and long-range antiferromagnetic (AFM) Heisenberg exchange interactions, plays a significant role in stabilizing chiral spin structures, such as the Yoshimori-type spin spiral state \cite{PhysRevB.111.L020405, 10.1063/5.0272907}. Moreover, magnetic anisotropy (MA) \cite{Lado_2017, PhysRevB.103.014438} also plays a crucial role in reshaping the energy landscape and determining the stability of magnetic phases and critical fields for phase transitions, such as skyrmion nucleation or annihilation. Crucially, the intricate interplay between DMI, exchange frustration, and MA plays a decisive role in determining the morphology and stability of chiral spin textures \cite{PhysRevB.107.054408, PhysRevB.109.214405, 2025arXiv250910661A, 2025arXiv250818522H, PhysRevB.111.L020405}. Consequently, controlling the competition between these distinct magnetic interactions in realistic material systems may induce variations in magnetic phases, which not only deepens the fundamental understanding of spin textures but also paves the way for prospective applications in next-generation nanodevices.

In this work, we take Li-absorbed Janus CrTeSe as an example to explore how the interplay among exchange frustration, DMI and MA can be manipulated to regulate chiral spin textures. Using a combination of first-principles calculations and atomic spin dynamics simulations, we examine the magnetic properties of the 2D Janus material CrTeSe functionalized by selective Li adsorption on either its Se- or Te-terminated surface. Our results reveal that the Li adsorption on different surfaces of Janus CrTeSe leads to different magnetic phases, such as Yoshimori spin spiral state and skyrmionic crystal state. A detailed analysis shows that this difference arises from the variation of interplay between exchange frustration, DMI and MA when Li is adsorbed onto different surfaces. Furthermore, we find that the evolution of magnetic textures under an external magnetic field differs significantly depending on the Li adsorption site. These findings demonstrate that surface adsorption engineering provides a promising route for reversible control of chiral magnetic textures in 2D materials. They also deepen the understanding of magnetic interactions and offer guidelines for designing tunable spintronic devices based on Janus magnets.

\section{\label{sec:level2}METHODS}
\subsection{First Principle Calculation}
All first-principles calculations are performed based on density functional theory (DFT) using the Vienna Ab initio Simulation Package (VASP) \cite{PhysRevB.47.558, PhysRevB.54.11169, KRESSE199615}. The exchange-correlation effect of electron-electron interactions is described by the Perdew-Burke-Ernzerhof (PBE) \cite{PhysRevB.44.13298, PhysRevB.59.1758} generalized gradient approximation (GGA) \cite{PhysRevLett.77.3865}, and the interaction between ionic cores and valence electrons is handled via the projector augmented-wave (PAW) method. Considering the strong correlation of electrons in the $3d$ orbitals of Cr atoms, a Hubbard $U$ \cite{PhysRevB.57.1505} correction ($U$ = 3 eV) is introduced to improve the accuracy of describing electron correlation effects \cite{Hou2022, Wang2024}. The plane-wave cutoff energy is uniformly set to 500 eV to ensure the convergence of energy calculations. Since the studied system is a 2D material, a 30 \AA\ vacuum layer is constructed along the direction perpendicular to the 2D plane (z direction) to avoid spurious interactions between adjacent layers under periodic boundary conditions. The convergence criterion for the electronic self-consistent cycle is set as $1\times 10^{-6}$ eV. Structural relaxation is performed using the conjugate gradient algorithm until the Hellmann-Feynman force acting on each atom is less than 0.002 eV/\AA, ensuring that the lattice structure reaches the minimum energy state. Integration in the Brillouin zone (BZ) is performed using a $\Gamma$-centered $9\times 9 \times 1$ k-point mesh. After structural relaxation, the lattice constant of the Li-adsorbed CrTeSe system with Li on the Se-terminated surface (denoted LiCrTeSe-1) is $a = b = 3.88$ \AA\ and that of the system with Li on the Te-terminated surface (denoted as LiCrSeTe-2) is $a = b = 3.99$ \AA. To verify the structural stability of the CrTeSe system after Li adsorption, the phonon dispersion spectrum is calculated using the PHONOPY package \cite{TOGO20151}.

\subsection{LLG Simulations}

\begin{figure*}
\centering
\includegraphics[height=7.5cm]{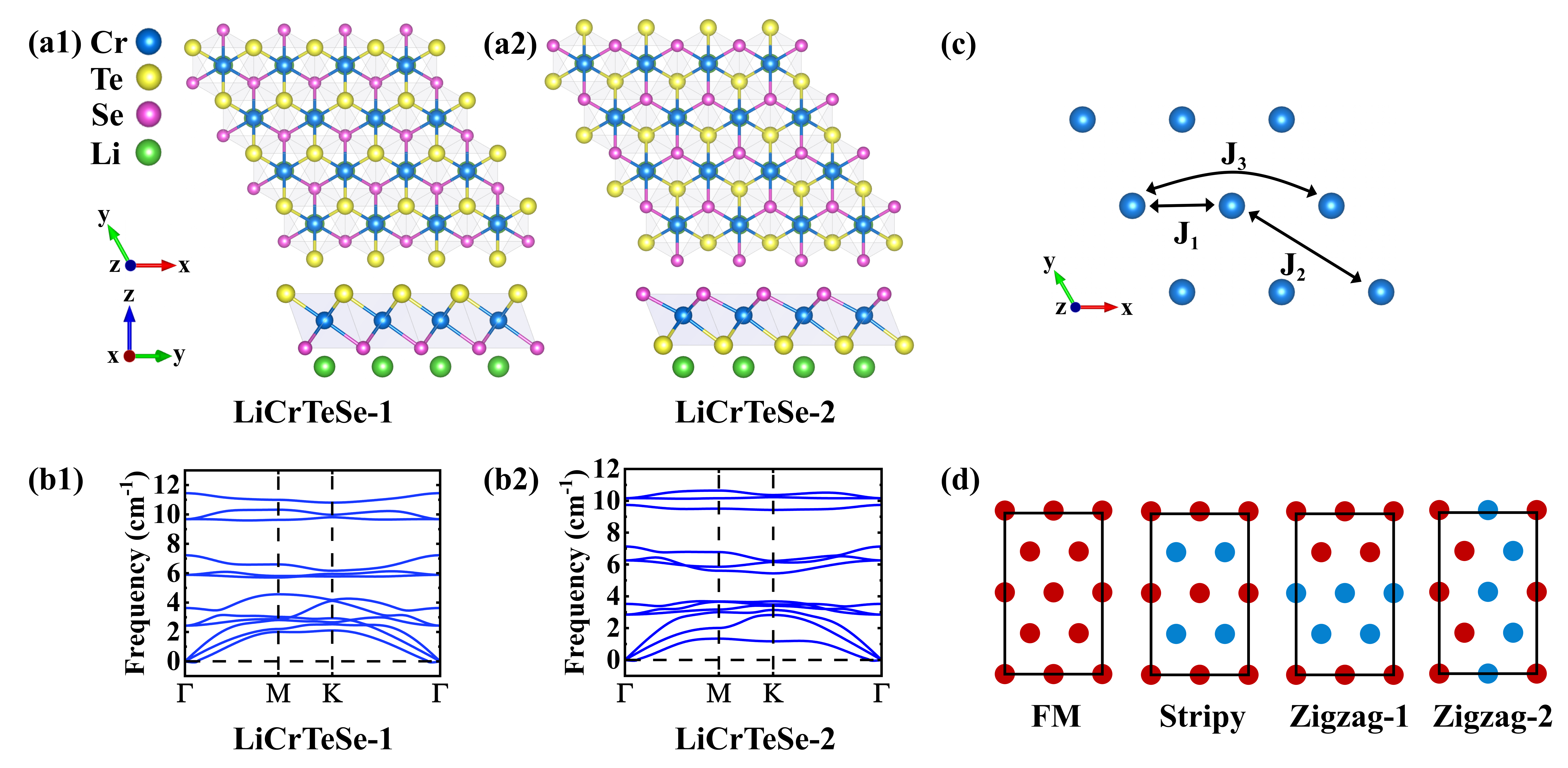}
\captionsetup{justification=raggedright,singlelinecheck=false}
\caption{\label{fig:figure1} 
(a1)-(a2) Crystal structures of LiCrTeSe-1 and LiCrTeSe-2, respectively. For each structure, the upper panel shows the top view along the $z$-axis, and the lower panel presents the side view perpendicular to the $z$-axis.
(b1)-(b2) Phonon spectra of LiCrTeSe-1 and LiCrTeSe-2, respectively.
(c) Schematic representation of Heisenberg exchange interactions, with first-nearest neighbor $J_1$, second-nearest neighbor $J_2$, and third-nearest neighbor $J_3$.
(d) Schematic of four distinct magnetic configurations constructed to calculate the Heisenberg exchange parameters. In the diagram, red spheres represent atoms in the spin-up state and blue spheres represent atoms in the spin-down state.
}
\end{figure*}

Atomic spin dynamic simulations are based on the Landau-Lifshitz-Gilbert (LLG) equation \cite{1353448} and implemented within the framework of the Spirit package \cite{PhysRevB.99.224414}. The simulation parameters are derived from the results of first-principles calculations in order to ensure the consistency of the theoretical framework. A $100\times 100$ 2D supercell based on the optimized hexagonal lattice, containing $10^4$ Cr atoms, is constructed to balance computational efficiency and the size effect of the system. The Gilbert damping coefficient in the LLG equation is set to $\alpha$ = 0.1, and the total number of iteration steps is $5\times 10^5$ to ensure that the spin configuration is fully relaxed to the energy minimum. The convergence criterion is set such that the total energy change of the system is less than $1\times 10^{-7}$ eV per atom and the change in the spin vector of each atom is less than $1\times 10^{-6}$. 

\section{\label{sec:level3}RESULTS AND DISCUSSION}
\subsection{Electronic and magnetic properties}
To facilitate discussion, we denote the Li-adsorbed structures on the Se- and Te-terminated surfaces as LiCrTeSe-1 and LiCrTeSe-2, respectively. The fully relaxed crystal structures of LiCrTeSe-1 and LiCrTeSe-2 are illustrated in Fig.~\ref{fig:figure1}(a1) and \ref{fig:figure1}(a2). The space group of the system is noncentrosymmetric $P3m1$ (No.156), which allows the presence of the DMI due to the lack of inversion symmetry. The Cr atomic layer forms a sandwich-like structure between the upper Se (or Te) atomic layer and lower Te (or Se) atomic layer. Each Cr atom is coordinated by three adjacent Se and three adjacent Te atoms to construct a distorted octahedral geometry, and the Cr atom is located at the center of the octahedron. Upon Li adsorption, the valence state of Cr changes from $+4$ to $+3$. The octahedral crystal field formed by Te and Se makes the $3d$ orbital of the $\mathrm{Cr}^{3+}$ ion approximately split into fully occupied threefold-degenerate $t_{2g}$ levels and empty twofold-degenerate $e_g$ levels in the octahedral coordinate system defined by the chalcogen ligands. In LiCrTeSe-1 (LiCrTeSe-2), the Cr-Se and Cr-Te bond lengths are 2.76 \AA\ (2.61 \AA) and 2.73 \AA\ (2.86 \AA), respectively. These slight differences in bond lengths result in octahedral distortion, which is more pronounced in LiCrTeSe-2. To further clarify the adsorption behavior of Li on Janus CrTeSe, we calculate the adsorption energies $E_a$ for LiCrTeSe-1 and LiCrTeSe-2 using the expression
\begin{eqnarray}
\begin{aligned}
    E_a = E_{\text{LCTS}}-E_{\text{Li}}-E_{\text{CTS}}
    \label{eq:adsorpsion}
\end{aligned}
\end{eqnarray}
where $E_{\text{LCTS}}$, $E_{\text{Li}}$ and $E_{\text{CTS}}$ are the energy of monolayer LiCrTeSe-1 (or LiCrTeSe-2), a single Li atom and the pristine CrTeSe, respectively. The obtained values are -2.61 eV/Cr for LiCrTeSe-1 and -2.07 eV/Cr for LiCrTeSe-2, implying that Li adsorption is energetically favored on the Se-terminated surface. The phonon spectra of the two structures, shown in Fig.~\ref{fig:figure1}(b1) and \ref{fig:figure1}(b2), are calculated using the finite displacement method. No imaginary frequencies are observed throughout the entire BZ, indicating that both structures are dynamically stable.

We construct an effective magnetic Hamiltonian for this 2D system:
\begin{eqnarray}
\begin{aligned}
    H=&J_1\sum_{\langle i,j\rangle}{\bm{S_i}}\cdot \bm{S_j}+
    J_2\sum_{\langle\langle i,j\rangle\rangle}\bm{S_i}\cdot \bm{S_j}+
    J_3\sum_{\langle\langle\langle i,j\rangle\rangle\rangle}\bm{S_i}\cdot \bm{S_j}\\
    &+A\sum_i (S_{iz})^2+
    \sum_{\langle i,j \rangle}\bm{D_{ij}}\cdot (\bm{S_i} \times \bm{S_j})
    \label{eq:Hamiltonian}
\end{aligned}
\end{eqnarray}
In this equation, $\bm{S_{i(j)}}$ denotes the normalized spin vector of the i-th (j-th) Cr atom. $S_{iz}$ denotes the $z$-component of the i-th spin. $J_1,J_2$ and $J_3$ correspond to the Heisenberg exchange parameters of the first, second, and third nearest neighbors, respectively, as shown in Fig.~\ref{fig:figure1}(c). $A$ is the single-ion MA energy. $\bm{D_{ij}}$ is the DMI between the i-th and j-th Cr atoms, and only the first nearest interaction is considered in this work. To achieve the Heisenberg exchange parameters, we construct four different magnetic configurations, which are shown in Fig.~\ref{fig:figure1}(d): fully ferromagnetic (FM), stripy antiferromagnetic (Stripy), and two zigzag antiferromagnetic configurations (Zigzag1 and Zigzag2). The energies of the four configurations are as follows:
\begin{eqnarray}
\begin{aligned}
    E_{\text{FM}}=E_0+24J_1|\mathbf{S}|^2+24J_2|\mathbf{S}|^2+24J_3|\mathbf{S}|^2,\\
    E_{\text{Stripy}}=E_0-8J_1|\mathbf{S}|^2-8J_2|\mathbf{S}|^2+24J_3|\mathbf{S}|^2,\\
    E_{\text{Zigzag-1/2}}=E_0\pm 8J_1|\mathbf{S}|^2 \mp 8J_2|\mathbf{S}|^2-8J_3|\mathbf{S}|^2
    \label{eq:exchange energy}
\end{aligned}
\end{eqnarray}
Here $E_0$ is the non-magnetic energy, and the spin vector $\mathbf{S}$ is normalized as 1. By solving the four equations, we obtain the values of $J_1, J_2$ and $J_3$ for LiCrTeSe-1 and LiCrTeSe-2, with the detailed derivation provided in the Supplemental Material \cite{supp}. The MA energy $A$ is determined by the energy difference between the in-plain FM state ($E_x$) and the out-of-plane FM state ($E_z$), following the formula:
\begin{figure*}
\centering
\includegraphics[height=3cm]{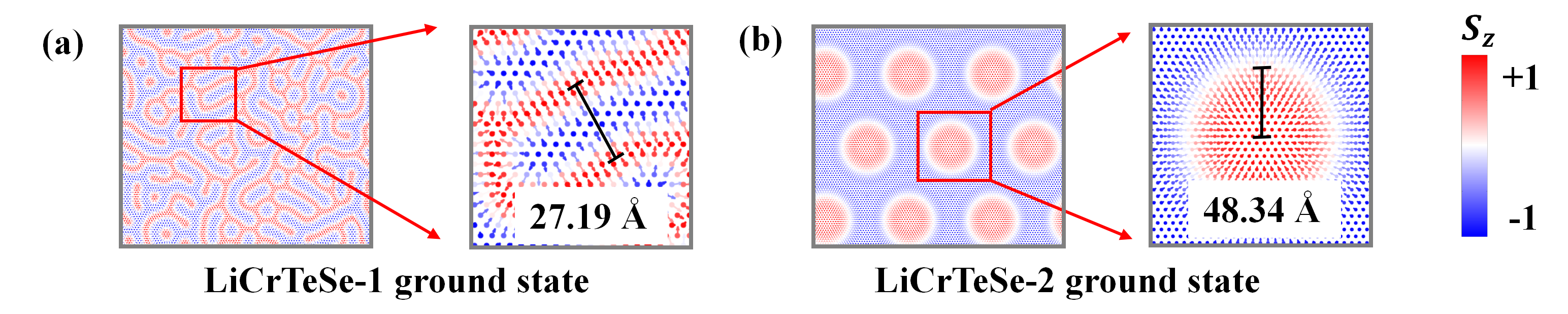}
\captionsetup{justification=raggedright,singlelinecheck=false}
\caption{\label{fig:figure2}The color map indicates the out-of-plane spin component.
(a) Magnetic ground states of LiCrTeSe-1. The left panel presents the overall magnetic configuration, while the right panel shows a magnified view of the red boxed region. This magnified view reveals a Yoshimori spin spiral with a period of 27.19 \AA\ (corresponding to seven lattice constants).
(b) Magnetic ground state of LiCrTeSe-2. The left panel displays the global magnetic structure, and the right panel provides a magnified view of the red boxed region. This enlarged view highlights an isolated skyrmion with a radius of 48.34 \AA\ (corresponding to twelve lattice constants).
}
\end{figure*}
\begin{eqnarray}
\begin{aligned}
    A=\frac{E_z-E_x}{|\mathbf{S}|^2}
    \label{eq:MAE}
\end{aligned}
\end{eqnarray}
A positive value of $A$ indicates in-plane magnetic anisotropy (IMA), while a negative value indicates out-of-plane magnetic anisotropy or perpendicular magnetic anisotropy (PMA). It should be noted that the MA effect originates from SOC, so the SOC is switched on during the calculation of $A$.

For DMI, according to the Moriya rule \cite{DZYALOSHINSKY1958241}, if a mirror plane passes through the midpoint of adjacent Cr atoms and is perpendicular to their bond, then the DMI vector $\bm{D}_{ij}$ for each pair of Cr atoms, which is perpendicular to their bond, can be expressed as Eq.~\eqref{eq:DMI-1}.
\begin{eqnarray}
\begin{aligned}
    \bm{D_{ij}}=d_{//}( \hat{u}_{ij}\times \hat{z})+d_z \hat{z}
    \label{eq:DMI-1}
\end{aligned}
\end{eqnarray}
where $\vec{u}_{ij}$ is the unit vector that points from $\bm{S}_i$ to $\bm{S}_j$, and $\vec{z}$ is the unit vector along the out-of-plane direction. Previous studies on 2D magnetic materials with a triangular lattice and $C_{3v}$ symmetry have shown that the out-of-plane component $d_z$ has a limited effect on the formation of spin spirals and skyrmions \cite{PhysRevB.109.214405}. Thus, only the in-plane component $d_{//}$ is considered in this work. $d_{//}$ is obtained from the energy difference between two chiral spin configurations (clockwise, $E_{\text{CW}}$, and counterclockwise, $E_{\text{ACW}}$) constructed in a $4\times 1$ supercell (see Fig. S4 in Supplemental Material \cite{supp}), as expressed in Eq.~\eqref{eq:DMI-2}.
\begin{eqnarray}
\begin{aligned}
    d_{//}=\frac{E_{\text{CW}}-E_{\text{ACW}}}{12|\mathbf{S}|^2}
    \label{eq:DMI-2}
\end{aligned}
\end{eqnarray}
The calculated magnetic parameters for the two configurations are shown in Table~\ref{tab:table1}. It can be observed that the nearest-neighbor exchange interaction $J_1$ in both systems is FM, while the third-nearest neighbor coupling $J_3$ is AFM, leading to exchange frustration and resulting in a variety of magnetic configurations. Furthermore, $J_3$ in LiCrTeSe-1 is significantly larger than in LiCrTeSe-2, indicating that the exchange frustration in LiCrTeSe-1 is much stronger. This enhancement of $J_3$ likely originates from stronger Cr-X (X = Te, Se) orbital hybridization and reduced Cr-Cr distance in LiCrTeSe-1, which together strengthen the AFM superexchange interaction (see Supplemental Material \cite{supp}). Notably, the MA energy of LiCrTeSe-1 is 2.15 meV/Cr, corresponding to IMA, whereas that of LiCrTeSe-2 is -0.80 meV/Cr, corresponding to PMA. The distinct MA behaviors originate from the different orbital hybridizations between the Te and Se $p$ orbitals near the Fermi level (see Supplemental Material \cite{supp}). Using the parameters from Table~\ref{tab:table1}, we perform LLG simulations for both configurations, and the magnetic ground states are shown in Fig.~\ref{fig:figure2}(a) and \ref{fig:figure2}(b). LiCrTeSe-1 stabilizes a spin spiral interspersed with a small number of skyrmions. Importantly, simulations with the DMI switched off (see Fig.~S5 in Supplemental Material \cite{supp}) confirm that the spiral remains, demonstrating that its origin lies in exchange frustration. Therefore, we can conclude that the ground state of LiCrTeSe-1 is a Yoshimori-type spiral stabilized by exchange frustration, with the DMI further selecting its chirality and defining rotational plane \cite{doi:10.1021/acs.nanolett.5c03355}. As shown in Fig.~\ref{fig:figure2}(a), the red box (magnified on the right) highlights the N$\Acute{\textup{e}}$el-type spiral structure, which has a period of approximately 27.19 \AA\ (corresponding to seven lattice constants). In contrast, LiCrTeSe-2 hosts only skyrmionic state without stripe domains. The magnified region in Fig.~\ref{fig:figure2}(b) shows a typical skyrmion with a radius of roughly 48.34 \AA\ (corresponding to twelve lattice constants). The difference of magnetic textures between LiCrTeSe-1 and 2 underscores the decisive role of Li adsorption in tuning the magnetic ground state.

\begin{table}
\captionsetup{justification=raggedright,singlelinecheck=false}
\caption{\label{tab:table1} Magnetic parameters of LiCrTeSe-1 and LiCrTeSe-2. 
For simplicity, spins have been normalized to $|\mathbf{S}|=1$. The energy unit is meV/Cr. One can refer to Eq.~\eqref{eq:Hamiltonian} and Eq.~\eqref{eq:MAE}-\eqref{eq:DMI-2} for specific meaning of the parameters.
}
\begin{ruledtabular}
\begin{tabular}{cccccc}
 & $J_1$ & $J_2$ & $J_3$ & $A$ & $d_{\mathop{//}}$ \\
\hline
LiCrTeSe-1 & -21.62 & -2.19 & 8.09 & 2.15 & 3.76 \\
LiCrTeSe-2 & -26.67 & -0.86 & 2.16 & -0.80 & 3.63 \\
\end{tabular}
\end{ruledtabular}
\end{table}

\begin{figure*}
\centering
\includegraphics[height=9cm]{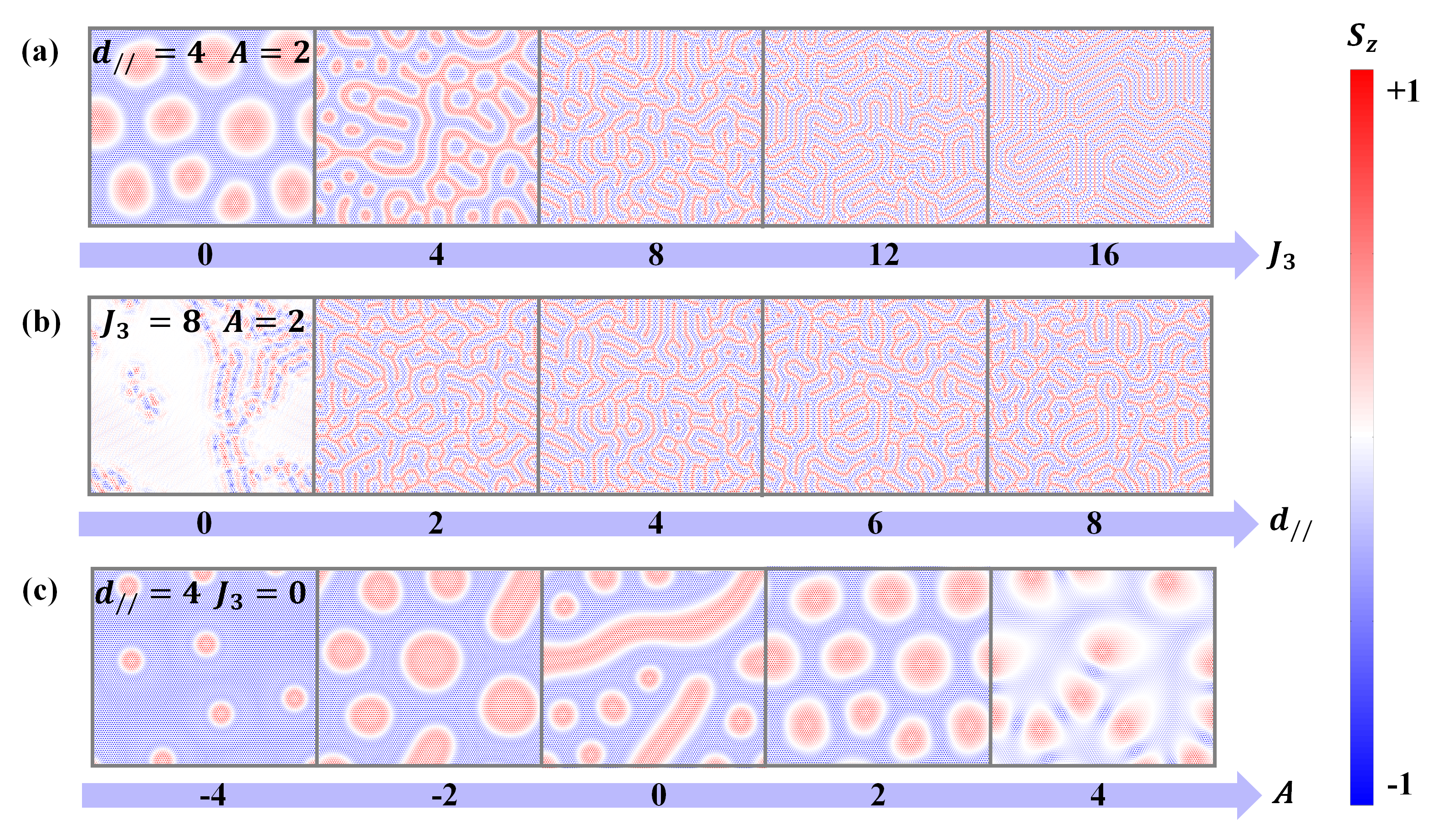}
\captionsetup{justification=raggedright,singlelinecheck=false}
\caption{\label{fig:figure3}
The color map indicates the out-of-plane spin component.
(a) Phase diagram of the toy model with parameters $J_1=-20$ meV/Cr (FM), $d_{//}=4$ meV/Cr, $A=2$ meV/Cr (IMA), and $J_2=0$, while tuning $J_3$ from 0 to 16 meV/Cr (AFM).
(b) Phase diagram of the toy model with parameters $J_1=-20$ meV/Cr (FM), $A=2$ meV/Cr (IMA), $J_2=0$, and $J_3=8$ meV/Cr (AFM), while varying $d_{//}$ from 0 to 8 meV/Cr.
(c) Phase diagram of the toy model obtained by fixing $J_1=-20$ meV/Cr (FM), $d_{//}=4$ meV/Cr, $J_2=J_3=0$, and varying the anisotropy $A$ from –4 meV/Cr (PMA) to +4 meV/Cr (IMA).
}
\end{figure*}

\subsection{Magnetic Phase Diagram}
An intriguing question naturally arises: why does Li adsorption on the Te-terminated and Se-terminated surfaces of Janus magnets lead to such distinct magnetic ground states? The most direct explanation lies in the delicate competition among the fundamental interactions in the spin Hamiltonian, namely, the exchange frustration, MA, and the DMI, all of which jointly regulate the stability of noncollinear spin textures. To disentangle their respective roles in stabilizing spin spirals and skyrmions, we construct a simplified toy model based on the triangular lattice and systematically vary the interaction parameters. The resulting phase diagrams are summarized in Fig.~\ref{fig:figure3}, with additional details provided in the Supplemental Material \cite{supp}.

To investigate the effect of $J_3$ on the magnetic configurations in LiCrTeSe-1, we fix $J_1=-20$ meV/Cr (FM), $d_{//}=4$ meV/Cr, $A=2$ meV/Cr (IMA), $J_2=0$, while tuning $J_3$ from 0 to 16 meV/Cr (AFM). As shown in Fig.~\ref{fig:figure3} (a), when $J_3=0$, skyrmions spontaneously emerge. This demonstrates that the interplay of $J_1$ and DMI is sufficient to stabilize skyrmions. Upon increasing $J_3$, mixed states appear containing both skyrmions and spin spirals. The skyrmion size shrinks, while the period of the spin spiral becomes shorter, resulting in denser stripe patterns. These results highlight the role of exchange frustration between $J_1$ and $J_3$ in favoring spiral states, with stronger frustration leading to shorter spiral periods and denser spin textures. This mechanism provides a key explanation for the distinct magnetic ground states observed in LiCrTeSe-1 and LiCrTeSe-2, as the much stronger $J_3$ in LiCrTeSe-1 enhances exchange frustration and favors a spiral state, whereas the weaker $J_3$ in LiCrTeSe-2 suppresses spiral formation, leading instead to a skyrmionic state.

Next, we examine the role of DMI by fixing $J_1=-20$ meV/Cr, $A=2$ meV/Cr, $J_2=0$, $J_3=8$ meV/Cr, while tuning $d_{//}$ from 0 to 8 meV/Cr, as shown in Fig.~\ref{fig:figure3}(b). At $d_{//}=0$, the ground state consists of bimerons embedded in an in-plane spiral background, demonstrating that exchange frustration alone is sufficient to stabilize spiral states even in the absence of DMI. In this regime, the competition between exchange interactions and MA favors in-plane spin rotations. Upon introducing DMI, however, the spin spiral evolves from an in-plane configuration into an out-of-plane N$\Acute{\textup{e}}$el-type spiral, with the DMI further selecting its chirality. This highlights the essential role of DMI in controlling both the orientation and handedness of spiral spin textures. With increasing $d_{//}$, mixed phases with coexisting skyrmions and spirals gradually emerge. 

We further assess the effect of MA by fixing $J_1=-20$ meV/Cr, $d_{//}=4$ meV/Cr, $J_2=J_3=0$, and varying $A$ from -4 meV/Cr (PMA) to +4 meV/Cr (IMA), as shown in Fig.~\ref{fig:figure3}(c). For strong IMA ($A=4$ meV/Cr), bimerons are stabilized, indicating that the interplay of DMI and strong IMA favors this texture. As $A$ shifts toward PMA, skyrmions reemerge with decreasing size and increasing isolation. This analysis reveals that MA plays a decisive role in determining the nature of chiral spin textures. In particular, the PMA intrinsic to LiCrTeSe-2, together with $J_1$ and DMI, stabilizes skyrmions rather than bimerons, consistent with the phase diagrams.

Taken together with the magnetic parameters summarized in Table~\ref{tab:table1}, these phase diagrams provide a comprehensive picture of how exchange frustration, MA, and DMI cooperate to determine the competition between Yoshimori-type spirals and skyrmion ground state. More detailed and systematic phase diagrams can be seen in Supplemental Material \cite{supp}.

\begin{figure*}
\centering
\includegraphics[height=6cm]{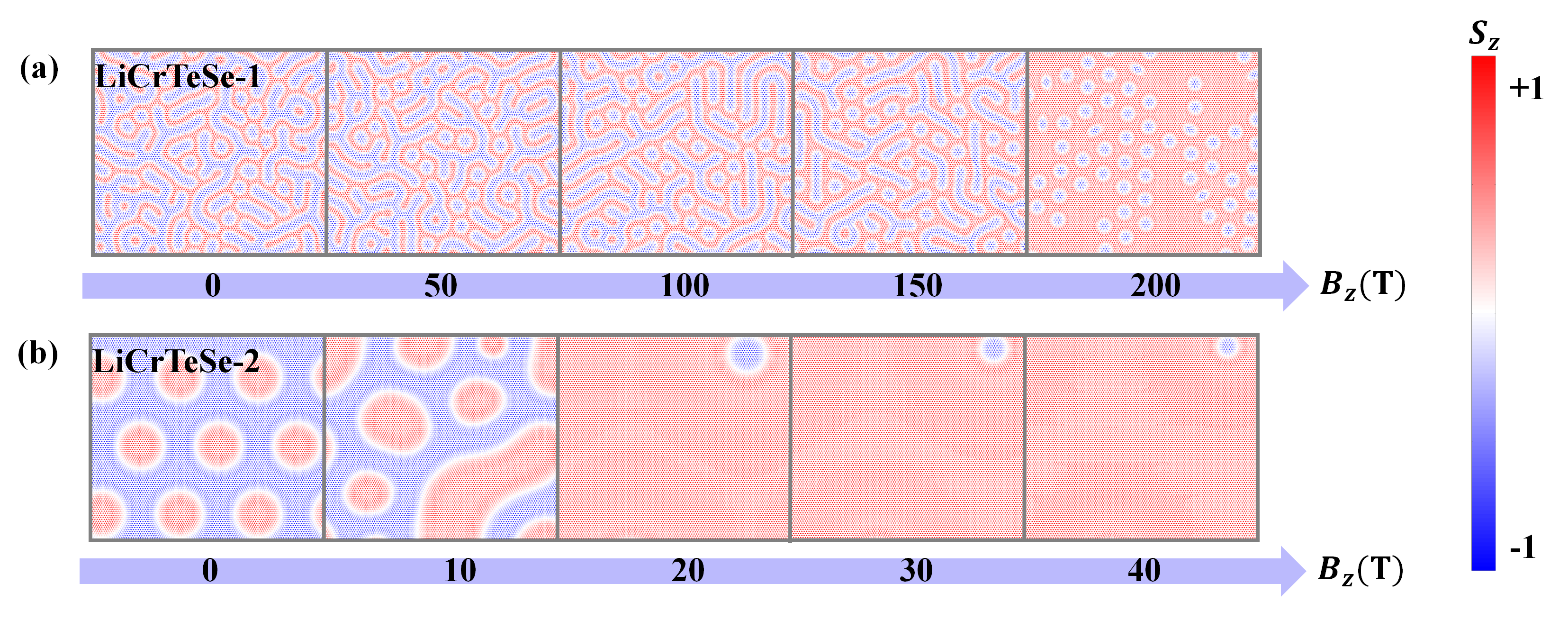}
\captionsetup{justification=raggedright,singlelinecheck=false}
\caption{\label{fig:figure4}
The color map indicates the out-of-plane spin component.
Magnetic ground state of (a) LiCrTeSe-1 and (b) LiCrTeSe-2 under an external out-of-plane magnetic field $B_z$.
}
\end{figure*}

\subsection{Magnetic textures under external magnetic field}
The difference among magnetic interactions in LiCrTeSe-1 and LiCrTeSe-2 would not only lead to distinct magnetic phases, but also give rise to different behaviors of magnetic textures under external field. To gain deeper insights, we examine the evolution of the magnetic ground states of LiCrTeSe-1 and LiCrTeSe-2 under an external out-of-plane magnetic field $B_z$. As shown in Fig.~\ref{fig:figure4}, both systems exhibit the general trend that increasing $B_z$ drives a transition into skyrmion states embedded in a FM background. With further enhancement of the field, the skyrmion size decreases and their density becomes progressively lower, reflecting the field-induced stabilization of compact, isolated skyrmions. A striking difference, however, lies in the critical field strength required to destabilize the initial ground states. In LiCrTeSe-1, the spiral phase remains robust up to $150$ T, gradually transforming into isolated skyrmions at around $200$ T, demonstrating the remarkable field stability of the spiral state. By contrast, in LiCrTeSe-2, isolated skyrmions emerge already at $20$ T, where the system evolves from a skyrmion lattice into a sparse array of individual skyrmions. These results reveal the dramatically different field sensitivities of chiral spin textures in LiCrTeSe-1 and LiCrTeSe-2.

To understand the different field-dependent behavior of spin textures in LiCrTeSe-1 and LiCrTeSe-2, we analyze the energy stability of the corresponding ground state spin textures with respect to field-polarized ferromagnetic state. To this end, we calculated the energy difference between the relevant magnetic ground states for the two systems and the out-of-plane ferromagnetic state, as shown in Table~\ref{tab:table2}. For LiCrTeSe-1, the ground state corresponds to a spin-spiral (SS) configuration, as shown in Fig.~\ref{fig:figure2}(a), whereas for LiCrTeSe-2 the ground state is a skyrmion (SK) phase, as shown in Fig.~\ref{fig:figure2}(b). The third row of each table reports the energy difference $\Delta E=E(\text{SS})-E(\text{FM})$ and $\Delta E=E(\text{SK})-E(\text{FM})$.

The results clearly reveal that the SS state is energetically more favorable than the FM state in both compounds, and the magnitude of $\Delta E$ directly reflects the field strength required to polarize the noncollinear ground state into out-of-plane FM state. Specifically, LiCrTeSe-1 and LiCrTeSe-2 exhibit total energy differences of -0.97 meV/Cr and -0.04 meV/Cr, respectively. Since an out-of-plane magnetic field ultimately favors the ferromagnetic alignment, a larger energy difference between the chiral magnetic configuration and the FM state implies a stronger field is required to induce the transition. Consequently, magnetic textures with a greater energy difference to the ferromagnetic state are less sensitive to external fields. Our results thus naturally explains why the magnetic ground state of LiCrTeSe-1 and LiCrTeSe-2 exhibit markedly different sensitivities to external magnetic fields.

\begin{table}[H]
\caption{\label{tab:table2} Upper: Total energy of spin-spiral (SS) state and FM state for LiCrTeSe-1, as well as the decomposed energies into each term of Hamiltonian. Energy unit is meV/Cr. The ground state corresponds to a spin-spiral (SS) configuration, as shown in Fig.~\ref{fig:figure2}(a).
Lower: Total energy of skyrmion state and FM state for LiCrTeSe-2, as well as the decomposed energies into each term of Hamiltonian. Energy unit is meV/Cr. The ground state corresponds to a skyrmion (SK) configuration, as shown in Fig.~\ref{fig:figure2}(b).
The third row of upper and lower sections reports the energy difference. The above calculations are performed in a $100 \times 100$ supercell containing $10^4$ Cr atoms.
}
\begin{ruledtabular}
\begin{tabular}{ccccccc}
 & Total & $J_1$ & $J_2$ & $J_3$ & $A$ & $d_{\mathop{//}}$ \\
\hline
SS state & -15.04 & -15.46 & -5.53 & -3.00 & -4.75 & -5.71  \\
FM state & -14.07 & -17.82 & -6.17 & 0 & -4.64 & -4.85 \\
$E(\text{SS})-E(\text{FM})$ & -0.97 & 2.36 & 0.64 & -3.00 & -0.11 & -0.86 \\
\hline\hline
SK state & -20.16 & -20.65 & -5.35 & -3.62 & -4.90 & -5.06  \\
FM state & -20.12 & -20.82 & -5.37 & -3.56 & -4.93 & -4.85\\
$E(\text{SK})-E(\text{FM})$ & -0.04 & 0.17 & 0.02 & -0.06 & 0.03 & -0.21 \\
\end{tabular}
\end{ruledtabular}
\end{table}

A more detailed breakdown reveals that in LiCrTeSe-1, the exchange contribution from $J_3$ provides a substantially larger energy gain in the chiral magnetic state compared to LiCrTeSe-2. This is consistent with the stronger third-nearest-neighbor Heisenberg exchange interaction in LiCrTeSe-1, which stabilizes the spiral ground state and renders it more resistant to field-driven transitions. In essence, the adsorption of Li on different sides of the Janus CrTeSe monolayer effectively tunes the strength of $J_3$, thereby modulating the system’s response to an external magnetic field.

\section{\label{sec:level4}CONCLUSION}
In summary, we combine first-principles calculations with LLG simulations to systematically investigate Janus CrTeSe with Li adsorption on opposite surfaces. We demonstrate that the adsorption site dictates the magnetic ground state: LiCrTeSe-1 (Li on Se side) favors a Yoshimori-type spin spiral, while LiCrTeSe-2 (Li on Te side) stabilizes skyrmions. This switching arises from adsorption-induced modifications in the interplay among exchange frustration, MA, and the DMI, which also lead to markedly different sensitivities of the two systems to external magnetic fields.

Overall, this work demonstrates that surface-selective Li adsorption regulates magnetic textures in Janus CrTeSe by tuning the interplay among exchange frustration, MA, and DMI. This provides a practical route for engineering spintronic devices with reversible switching between skyrmions and spin spirals. Moreover, recent experiments on Ir(110) have demonstrated that adsorption–desorption can drive such transformations \cite{doi:10.1021/acs.nanolett.5c03355}, highlighting the experimental feasibility of our proposed mechanism.

\begin{acknowledgements}

This work is supported by National Key R\&D Program of China (2023YFA1406400).

\end{acknowledgements}

%\bibliography{references}

%

\end{document}